\documentclass[superscriptaddress,secnumarabic,
amssymb,amsmath,nobibnotes,aps,prd,showkeys,showpacs,nofootinbib]{revtex4}%
\usepackage{graphicx}
\usepackage{epsf}
\usepackage{bm}
\usepackage{amsmath}
\usepackage{amsfonts}
\usepackage{amssymb}
\usepackage{epstopdf}
\usepackage{natbib}%
\providecommand{\U}[1]{\protect\rule{.1in}{.1in}}
\newcommand{\be}{\begin{equation}}
\newcommand{\ee}{\end{equation}}

\newcommand{\mincir}{\raise
-3.truept\hbox{\rlap{\hbox{$\sim$}}\raise4.truept\hbox{$<$}\ }}
\newcommand{\magcir}{\raise
-3.truept\hbox{\rlap{\hbox{$\sim$}}\raise4.truept\hbox{$>$}\ }}

\usepackage{color}

\begin{document}
\title{Gravitationally induced adiabatic particle production: From Big Bang to de Sitter}
\author{Jaume de Haro}
\email{jaime.haro@upc.edu}
\affiliation{Departament de Matem\`atica Aplicada I, Universitat Polit\`ecnica de Catalunya, Diagonal 647, 08028 Barcelona, Spain}
\author{Supriya Pan}
\email{span@iiserkol.ac.in}
\affiliation{Department of Physical Sciences, Indian Institute of Science Education and Research -- Kolkata, Mohanpur -- 741246, West Bengal, India}
\keywords{Dark energy, Particle creation, Non-equilibrium thermodynamics}
\pacs{98.80.-k, 05.70.Ln, 04.40.Nr, 98.80.Cq.}
\begin{abstract}
In the background of a flat homogeneous and isotropic space-time, we consider a scenario of the universe driven by
the gravitationally induced `adiabatic' particle production with constant creation rate. We have shown that this universe attains a big bang singularity
in the past and at late-time it asymptotically becomes de Sitter. To clarify this model universe, we perform a dynamical analysis and found that the universe attains a thermodynamic equilibrium in this late de Sitter phase. Finally, for the first time, we have
discussed the possible effects of `adiabatic' particle creations in the context of Loop Quantum Cosmology.

\end{abstract}

\maketitle
\section{Introduction}
{{}Late-time acceleration of our universe \cite{Perlmutter99, Riess98, Spergel2003,
Tegmark2004, Eisenstein2005, Komatsu2011} has become one of the fundamental puzzles
in modern cosmology. To explain this accelerating phase of the universe, mainly two known
approaches are used. One is the introduction of some component called dark energy
having large negative pressure with its equation of state ``$\omega< -1/3$'' \cite{CST}, and the other one is
the modifications in the standard General Relativity \cite{odintsov}. As a result, various
dark energy models, as well as, modified gravity models have been introduced
in order to describe the current accelerating universe. However, among such
various modifications both in matter, as well as, in gravity sectors, $\Lambda$CDM cosmology is supported by a large number of observational data. Still,
we are worried about two biggest and famous problems in $\Lambda$, the cosmological
constant problem \cite{Weinberg1989} and the coincidence problem \cite{Steinhardt2003}. Additionally,  it has been recently pointed out that our universe has slight phantom nature \cite{Planck2014, Rest2014, XLZ2013, CH2014, SH2014}, i.e. the equation of state of such component driving the cosmic acceleration goes beyond ``$-1$'' (For a comprehensive discussion on phantom cosmology we refer \cite{GWG}).
Hence, people have been trying to find some
other way (s) so that we can realize an actual description for our universe in agreement with the latest astronomical data we have.\newline

On the other hand, besides the above two known approaches, namely, the dark energy
and the modified gravity theories, recently, considerable attention has been given to the
gravitationally induced particle production. The gravitational particle production has a long history in the cosmological domain. Originally, the microscopic description of particle production follows from a seminal paper by Schr\"{o}dinger \cite{Schrodinger} in 1939. Later on, Parker and collaborators \cite{Parker} re-investigated this microscopic particle production by the gravitational field of an expanding universe based on the Bogoliubov mode-mixing technique in the context of quantum field theory in a curved space-time \cite{Birrell}. In spite of being diligent and well motivated, the above microscopic description of particle creation was not fully recognized in the cosmological context since the absence of a proper methodology in order to connect them with the classical Einstein's field equations. However, soon after some years, Prigogine et al \cite{Prigogine98} studied the macroscopic description of the particle production based on the non-equilibrium thermodynamics of open systems and described how to insert this particle creation mechanism into the Einstein's field equations in a consistent way. Just after that, Calv\~{a}o, Lima and Waga \cite{CLW92} re-discussed this macroscopic description of particle production by formulating a covariant approach, where the back reaction term is naturally included into the Einstein's field equations whose negative pressure could provide a self-sustained mechanism for current cosmic acceleration. Yet this description is still incomplete in the sense that the particle creation rate should be calculated from the quantum field theory in curved spacetime \cite{Birrell}. We note that there is a difference between the matter creation by the gravitational field of an expanding universe and the mechanism of bulk viscosity which had been proposed earlier by Zeldovich \cite{Zeldovich} to account for the particle production. This difference has been discussed by Lima and Germano \cite{LG92} showing that although both the mechanisms can depict the same cosmic evolution, but both the processes are completely different from the thermodynamical view point. However,  in this connection, we mention that although there is an analogy between the present matter creation models and the models developed by Hoyle and Narlikar \cite{HN} (known as Steady State Cosmology) adding extra terms to the Einstein-Hilbert action interpreting the so-called C-filed, but they are completely different in the sense that, in the later case, the creation phenomenon is interpreted through a process of interchanging of energy and momentum between matter itself and the C-field.\newline

Recently, accelerating cosmology driven by the gravitationally induced ``adiabatic'' (entropy per particle remains constant during the process) particle production has intensively been examined in the FLRW universe \cite{SSL09, LJO10, LB10, LBC12, JOBL11, LGPB14, CPS14, NP15, FPP14, NP2016, RCN2016}. Further, it has also been discussed that not only the current accelerating universe, the particle production can also take into account for the early inflationary universe \cite{Inflation}. In fact, it has been shown that the matter creation models can provide an alternative cosmology known as CCDM cosmology (CCDM $\equiv$ creation cold dark matter) \cite{SSL09, LJO10, LB10}  which is a viable alternative to the $\Lambda$CDM model both at background and perturbative levels \cite{JOBL11}. Additionally, the effects of adiabatic particle production have been tested in the cosmic microwave background level \cite{NP2016} which shows a close behavior to that of $\Lambda$CDM. In fact, it has been argued that adiabatic particle production can provide a possible connection between the early and late accelerating regimes \cite{RCN2016}. Further, it has been argued that a complete cosmic scenario with early and late de Sitter eras can also be encountered by such mechanism \cite{LBC12}. The stability of such models in agreement with the generalized second law of thermodynamics have been studied \cite{MP13,LGPB14}. However, the key point in all such models driven by the gravitational particle production is to consider several choices for the rate of particle production, which in general are considered to be the functions of the Hubble rate of the FLRW universe. Along with several choices for this creation rate, the possibility of constant creation rate was also considered to understand the dynamics of the universe in the pioneer work of Prigogine et al \cite{Prigogine98} (see Section \ref{sec:bigbang} of the present work) and recently discussed in \cite{Chakraborty14}, where the author
concludes, in a wrong way that ``adiabatic'' particle production with a constant creation rate leads to an emergent universe. However, the crucial point is that since the universe evolves from a high energy scale to the present low energy regime then the question remains on how the particle creation could be constant irrespective of the energy scales, rather it should depend on some energy scale. But, as this constant particle creation rate is associated with some critical issues in the literature \cite{Prigogine98,Chakraborty14}, so keeping this problem in mind as well as to clarify these issues related to this constant rate, in the present work, we consider our universe modelled by
some constant particle production rate in addition to a perfect fluid satisfying the barotropic equation of state.
We realized some interesting possibilities. 
First of all, we found the analytic solutions for Hubble parameter and the scale factor
which readily show that, initially, the universe must have realized the big bang singularity, and then asymptotically it reaches to the de Sitter phase. The results
from dynamical analysis perspective assure that the scenario from the big bang to de Sitter
is physically viable. We extend, for the first time, the notion of particle production to the Loop Quantum Cosmology,
where holonomy corrections introduce a quadratic term in the Friedmann equation (see \cite{singh} for details)
which results in a non-singular bouncing cosmology, in which the big bang singularity is replaced
by a non-singular bounce \cite{Singh, bho}, that ends in the expanding de Sitter regime in an asymptotic manner. However,  we note that, in this formalism the replacemnt of the big bang singularity by such a nonsingular bounce is only realized if the adiabatic creation rate is considered as a function of some energy scale.

\vspace{0.5cm}

{The paper is organized as follows: In Section \ref{sec:ede}, we consider the universe as an open thermodynamical system, where we obtain the corresponding dynamical equations and show that the particle
production acts as an effective dark energy.  In section \ref{sec:dynamical}, we deal with our model as a dynamical system, in this way, we see that at late times, there is an attractor  de Sitter phase which could
explain the current acceleration of the universe. section \ref{sec:bigbang} is devoted to the study of the big bang singularity in open systems revealing, in contrary to some earlier statements,
that it survives in that model. Later, a thermodynamic description of the current model has been discussed in section \ref{sec:thermo}. Further, in section \ref{sec:lqg}, we extend our model to Loop Quantum Cosmology, where the big bang singularity is replaced by a big bounce,
obtaining a bouncing universe that starts in the contracting phase
and after bouncing, it ends at late times in a de Sitter regime in the expanding phase. The model also be able to depict the so-called Matter Bounce Scenario (see for a recent review
\cite{hc}), in fact we have calculated
the spectral index and its running of cosmological perturbations obtaining theoretical values that fit well with observational data \cite{Ade}. Finally, in the last section \ref{discu}, we have summarized our results.

\vspace{0.5cm}
The units used throughout the paper are: $c=\hbar=8\pi G=1$.

\section{Effective Dark Energy}
\label{sec:ede}

We know that at large scales our universe is well described by the flat{{} Friedmann-Lema{\^\i}tre-Robertson-Walker (FLRW)} line element given by

\begin{equation}
ds^2= -dt^2+a^2(t) [dx^2+ dy^2+ dz^2],\label{FLRW}
\end{equation}
where $a(t)$ is the scale factor of the universe. 
Let us consider a closed physical volume $V$ containing
$N$ number of particles. Hence, the first law of thermodynamics  for closed systems, i.e, when there is not particle creation (or, equivalently,
the conservation of the internal energy $E$) states
\begin{equation}
dE= dQ-pdV,\label{1st-law}
\end{equation}
where $dQ$ is the amount of heat received by the system in time $dt$ and $p$ is the thermodynamic pressure. Now, unlike a closed thermodynamical system, an open thermodynamical system is much more reasonable where the particle numbers are not fixed as considered in \cite{Prigogine98}. Thus, when $N$ is variable, that means while dealing with an open system, where the number of particles is not conserved, the equation  (\ref{1st-law}) must be replaced by
\cite{Prigogine98}
\begin{equation}
dE= dQ-pdV +\left(\frac{p+\rho}{n}\right) dN ,\label{general-eqn}
\end{equation}
where  $n= N/V$ is the particle density and $\rho= E/V$ is the energy density.
By introducing the heat per unit particle $d\tilde{q}= dQ/N$, the above equation (\ref{general-eqn}) reduces to
\begin{equation}
d\left(\frac{\rho}{n}\right)= d\tilde{q}- p d\left(\frac{1}{n}\right),\label{general-simplified}
\end{equation}
which is known as the Gibb's equation.
For open systems, the non-conservation of the particle
number is written as ($N^{\mu} _{;\mu} \neq 0$) \cite{Prigogine98}
\begin{equation}
\dot{n}+\Theta n= n \Gamma~~~\Longleftrightarrow~~~ \dot{N}= \Gamma N \label{non-conservation}
\end{equation}
where $\Gamma$ stands for the rate of change of the particle number in a physical volume $V= V_0 a^3$ (where $V_0$ is the co-moving volume), $N^{\mu}= n u^{\mu}$ represents particle
flow vector, $u^{\mu}$ is the usual particle four velocity. For the FLRW universe, $\Theta= u^{\mu} _{;\mu}= 3 H = 3\, \dot{a} (t)/a(t)$, denotes the fluid expansion and by notation $\dot{n}= n_{,\mu} u^{\mu}$.
The positivity of $\Gamma$ means that, there is creation of particles while there will be annihilation of particles if $\Gamma$ is negative. However, in any case, non-zero $\Gamma$ will
introduce a dissipative effect (bulk viscous pressure) to the thermodynamical fluid and one has to take care of non-equilibrium thermodynamics.\newline

The energy conservation equation of the system follows from (\ref{general-eqn}) is
\begin{eqnarray}
\dot{\rho}&=& - 3H \left(1-\frac{\Gamma}{3H}\right) (p+ \rho)+ n T \dot{s},\label{energy-conservation}
\end{eqnarray}
where `$s$' is the specific entropy (entropy per particle, {{} i.e., $ds=\frac{d\tilde q}{T}$}) of the system, and $T$ is the fluid temperature.
We note that, the validity of the generalized second law of thermodynamics always demand $\Gamma \geq 0$. From Eq. (\ref{energy-conservation}), it is seen that the standard energy conservation equation can be recovered under two conditions: $|nT\dot{s}| \longrightarrow 0$, and $\Gamma/3H \ll 1$. Special attention has been given to the simplest cases where the entropy per particle remains constant, that means $\dot{s}= 0$, which are also referred to as `adiabatic' or `isentropic' condition. Thus,
under the isentropic condition, the conservation equation (\ref{energy-conservation}) takes the form \cite{SSL09, LJO10, LB10, LBC12, JOBL11, LGPB14, CPS14, NP15, FPP14, NP2016, RCN2016}
\begin{eqnarray}
\dot{\rho}&=& - 3H \left(1-\frac{\Gamma}{3H}\right) (p+ \rho),\label{energy-conservation1}
\end{eqnarray}
Since the adiabatic condition is the simplest phenomenon and also potential in the sense that it could account for the current accelerating universe \cite{SSL09, LJO10, LB10, LBC12, JOBL11, LGPB14, CPS14, NP15, FPP14, NP2016, RCN2016}, hence, in this work, we consider the cosmology of constant particle creation in an adiabatic manner and its consequences in the FLRW universe. Further, we can write down the Friedmann's equation and Raychaudhuri's equation as follows (recall in our units $8 \pi G= 1$):

\begin{equation}
H^2= \frac{\rho}{3},~~~~(\mbox{Friedmann's equation})\label{friedmann1}
\end{equation}
and

\begin{equation}
\dot{H}= -\frac{1}{2} \left(1-\frac{\Gamma}{3H}\right) (p+ \rho).~~~~(\mbox{Raychaudhuri's equation})\label{friedmann2}
\end{equation}
It should be noted that, any two of the three equations (\ref{energy-conservation1}), (\ref{friedmann1}), (\ref{friedmann2}) are independent. Therefore, in order to understand the evolution of the universe, a relation between $p$, and $\rho$, as well as,  $\Gamma$ should be prescribed. In principle, the functional form of particle creation rate $\Gamma$ should be decided from the quantum field theory (QFT) in curved space times where the particle creation process happens in an irreversible thermodynamic way, which awaits for a proper development of the QFT in curved spacetime.
However, one may notice that, the  late de Sitter expansion ($\equiv$ $\dot{H}= 0$, that means $H$ becomes constant) is realized for $\Gamma=$ constant. We consider that the above cosmic substratum be a perfect fluid with barotropic  Equation of State (EoS): $p= (\gamma-1)\rho$, where $\gamma$ is a constant
satisfying $\gamma \geq 0$. Also, we consider that, $\Gamma$ to be constant for our whole analysis. Now, due to the adiabatic particle creations, the effective equation of state $w_{eff}$ of the system takes
the form 
\begin{eqnarray}\label{EoS}
w_{eff}&\equiv&-1-\frac{\dot{\rho}}{3H\rho}= -1-\frac{2\dot{H}}{3H^2}= -1+ \gamma \left(1-\frac{\Gamma}{3H}\right),\label{eff-eos}
\end{eqnarray}
which leads us the following observations:\\

$\bullet$ For $H \gg \Gamma$, one has $w_{eff}(H) \cong -1+ \gamma$,  non-phantom domination. \\

$\bullet$ {{} For $3 H \gtrsim\Gamma$, one has $w_{eff}(H) \gtrsim -1$},  accelerated expansion.\\

$\bullet$ For $3 H < \Gamma$, one has $w_{eff}(H) < -1$,  phantom domination.\\

For constant $\Gamma$, the differentiation of (\ref{eff-eos}) with respect to the cosmic time $t$, and using Raychaudhuri's equation (\ref{friedmann2}), we find
{{}
\begin{eqnarray}
\dot{w}_{eff} (H) &=& -\frac{\gamma^2 \Gamma}{2} \left(1-\frac{\Gamma}{3H}\right)
,\label{dot-eos}
\end{eqnarray}}
from which we have the following observations:

I. $\dot{w}_{eff}< 0$, for $\Gamma< 3H$. Hence, this implies that, $w_{eff}$ deceases as $t$ increases.\\

II. $\dot{w}_{eff}> 0$, for $\Gamma> 3H$. Hence, this implies that, $w_{eff}$ increases as $t$ increases.\\

Finally, as we will show in Section \ref{sec:dynamical}, this result means that,
open systems could be understood as an effective kind of dark energy because, at late times, the universe will go asymptotically to a de Sitter phase ($w_{eff}\rightarrow -1$), which
is a plausible explanation of the current acceleration of the universe.

\section{Appearance of the Big Bang singularity}
\label{sec:bigbang}

Combining the Friedmnann and Raychaudhuri equations  (\ref{friedmann1}) and (\ref{friedmann2}), for the EoS $p=(\gamma-1)\rho$ one obtains
\begin{eqnarray}
	\dot{H}= -\frac{3\gamma}{2} \left(1-\frac{\Gamma}{3H}\right)H^2, \label{ray}
\end{eqnarray}
Now, Eq. (\ref{ray}) can be integrated as
\begin{eqnarray}
	H (t)&=& \frac{\Gamma }{3} \Bigg[\frac{\frac{H_0}{H_0-\frac{\Gamma}{3}}
		\exp\left(\frac{\Gamma \gamma}{2} (t-t_0)\right)}{\frac{H_0}{H_0-\frac{\Gamma}{3}}\exp\left(\frac{\Gamma\gamma}{2} (t-t_0)\right)-1}\Bigg],\label{Hubble-sol}
\end{eqnarray}
where $t_0$, $H_0$ are respectively the present day values of cosmic time, and the Hubble parameter. Now, we find in (\ref{Hubble-sol}) that, $H(t)$ becomes singular at some
finite time $t_s$, i.e., when
we have
\begin{eqnarray}
	\frac{H_0}{H_0-\frac{\Gamma}{3}}\exp\left(\frac{\Gamma \gamma}{2} (t-t_0)\right)-1&=& 0
\end{eqnarray}

Hence, the solution for $H (t)$ becomes
\begin{eqnarray}
	H (t)&=& \frac{\Gamma}{3} \Bigg[\frac{\exp\left({\frac{\Gamma\gamma}{2}(t-t_s)}\right)}{\exp\left({\frac{\Gamma\gamma}{2}(t-t_s)}\right)-1}\Bigg] \label{Hubble-sing},
\end{eqnarray}
and consequently, we find that
\begin{eqnarray}
	\lim_{t \rightarrow t_s^+} H (t)&=& \infty~~~~~~(\mbox{Big Bang singularity}).
\end{eqnarray}

Note also that, from formula (\ref{Hubble-sing}) we can calculate the age of the universe in this model, giving as a result
\begin{eqnarray}
	t_0-t_s=\frac{2}{\Gamma\gamma}\ln\left(\frac{H_0}{H_0-\frac{\Gamma}{3}}\right).
\end{eqnarray}

{{} In general,
	for} isentropic systems, the conservation equation {{} (\ref{general-eqn})} {{} could be written  as} \cite{Prigogine98}
{{}
	\begin{eqnarray}
		\dot{\rho}&=& \left(- 3H + \frac{\dot{N}}{N}\right)(p+ \rho),\label{energy-conservation3}
	\end{eqnarray}}
	where for the time being we consider that the cosmic substratum is a pressureless perfect fluid ($p= 0$, i.e., $\gamma= 1$) as considered in \cite{Prigogine98}. Also, we assume the
	simple relation $\rho=MN/V$ ($V= a^3$ is the volume of the FLRW universe), and further,
	we restrict us to the simple choice \cite{Prigogine98}
	{{}
		\begin{eqnarray}
			\dot{N}&=& \alpha V H^2=\frac{\alpha MN}{3},~~~\mbox{where}~\alpha \geq 0~.
		\end{eqnarray}
		which,  taking $\Gamma=\frac{\alpha M}{3}$, coincides with equation (\ref{non-conservation})}.
	Hence, replacing $\Gamma$ by $\frac{\alpha M}{3}$ in Eq. (\ref{Hubble-sing}), one gets
	{{}
		\begin{eqnarray}
			H (t)&=& \frac{M \alpha}{9} \Bigg[\frac{\exp\left({\frac{M\alpha}{6}(t-t_s)}\right)}{\exp\left({\frac{M\alpha}{6}(t-t_s)}\right)-1}\Bigg]\Longrightarrow
			a(t)=a_0\Bigg[\frac{\exp\left({\frac{M\alpha}{6}(t-t_s)}\right)-1}{\exp\left({\frac{M\alpha}{6}(t_0-t_s)}\right)-1}\Bigg]^{\frac{2}{3}}
			\label{Hubble-solution}
		\end{eqnarray}}
		and consequently, we find that
		
		\begin{eqnarray}
			\lim_{t \rightarrow t_s^+} H (t)= \infty \quad \mbox{and} \quad \lim_{t \rightarrow t_s^+} a (t)= 0,
		\end{eqnarray}
		which  shows that, there is a big bang singularity at $t= t_s$, and it falsifies the conclusion in \cite{Prigogine98}, where the authors claimed that, there is no big bang singularity
		in this formalism. On the other hand, we also find that
		\begin{eqnarray}
			\lim_{t \rightarrow \infty} H (t)&=& \frac{M \alpha}{9}~~~~~~~(\mbox{Asymptotically de Sitter})
		\end{eqnarray}

\section{Dynamical Analysis}
\label{sec:dynamical}

{{} To understand this system better, we need to calculate the critical points or the fixed points of the system (\ref{friedmann1}, \ref{friedmann2}). Note that, the
Friedmann equation (\ref{friedmann1}) is nothing but a constrain in General Relativity, which tells us that, the universe should follow a parabolic path in the plane ($H$, $\rho$).
However, solving for $\dot{H}= 0$, we find that, the above system (\ref{friedmann1}, \ref{friedmann2}) has two critical
points {$(0,~0)$, $\left(\frac{\Gamma}{3},~\frac{\Gamma^2}{3}\right)$}.\\

{{} Now, depending on the possible movements of the universe towards the non-zero critical point along the parabolic path described by the constrain (\ref{friedmann1}),
we have the following two scenarios.  Either the universe moves towards the critical point $\left(\frac{\Gamma}{3},~\frac{\Gamma^2}{3}\right)$ in the upward direction of the
parabola as shown in figure 1, or, it reverses its direction of movement, i.e., from $(0,0)$ to $\left(\frac{\Gamma}{3},~\frac{\Gamma^2}{3}\right)$.} For such
a scenario presented in figure 1, there must have some phantom fluid which drives the universe. In fact,  at  $(0,0)$ the universe starts to climb up the parabola, which is clearly
un-physical due to the nonexistence of no radiation and matter dominated eras.\\

Now, when the universe moves downward to the parabola as in figure 1, the scenario predicts a universe with an initial big bang singularity (for $H \gg \Gamma$, one may
have $\dot{H}= -\frac{3 \gamma}{2}H^2$, which is well-known that leads to a big bang.), and at late time, asymptotically it goes to the de Sitter phase, i.e., $H= \Gamma/3$, where to depict the current cosmic acceleration one has to
choose $\Gamma\sim H_0$, where $H_0$ is the current value of the Hubble parameter. In this case, the universe
pegged at $w_{eff}> -1$, for all time, and at late time, as the universe is almost de Sitter in nature, we have $w_{eff}(H) \gtrsim -1$.}\newline

In fact, if one chooses
$\gamma=\frac{4}{3}$, that is, we consider a radiation dominated universe ($p=\frac{1}{3}\rho$), then since $w_{eff}=\frac{1}{3}-\frac{4\Gamma}{9H}$, at early times ($H\gg \Gamma$) the
universe is radiation dominated, later when the Hubble parameter is close to $\frac{4\Gamma}{3}$, the universe enters in a matter dominated era, and finally at very late time
($H \gtrsim\frac{\Gamma}{3}$), the universe goes asymptotically to the de Sitter regime.\newline

\begin{figure}[h]
\begin{center}
\includegraphics[scale=0.40]{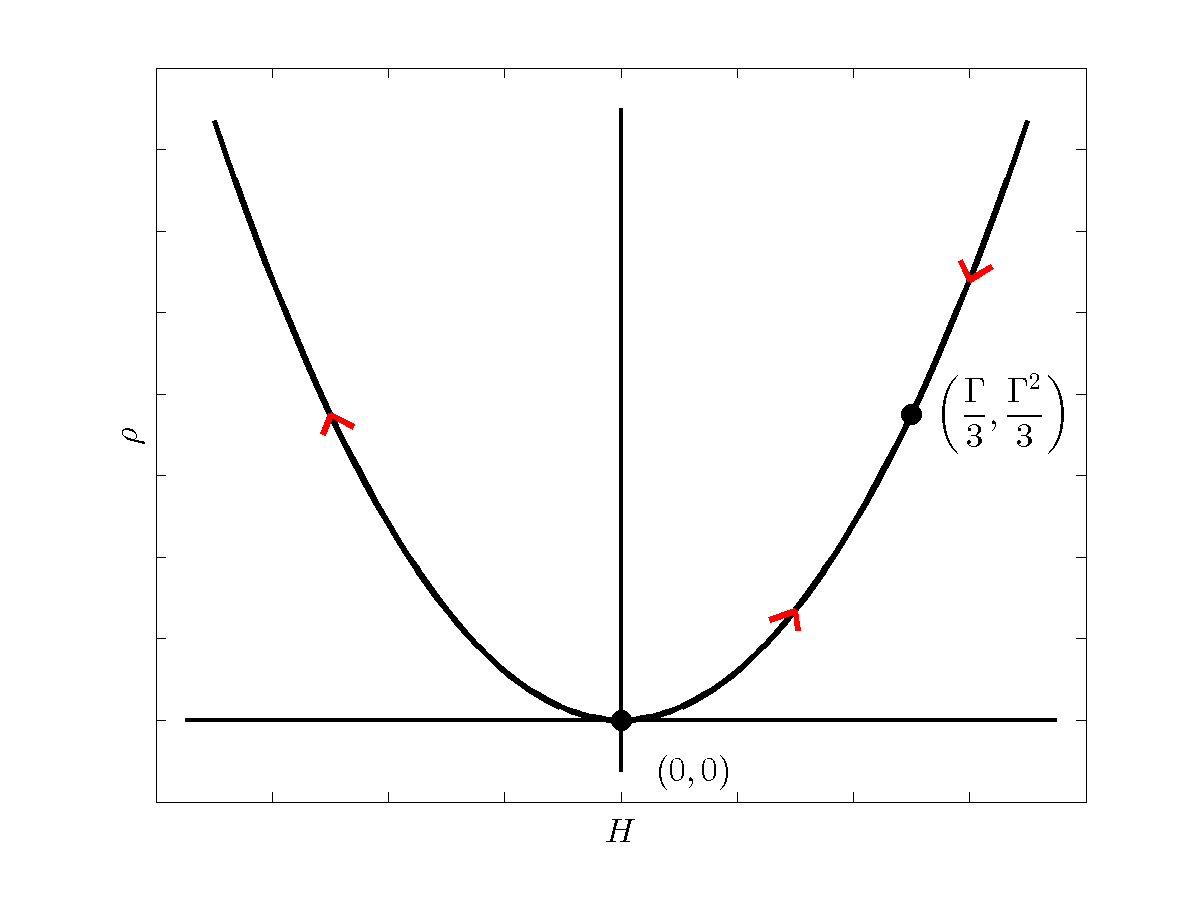}
\end{center}

\caption{{\protect\small Three different dynamics: One in the contracting phase starting at $(0,0)$, another that goes from $(0,0)$ to $(\frac{\Gamma}{3},\frac{\Gamma^2}{3})$ driven by a phantom fluid, and finally the dynamics driven by a non-phantom fluid which starts in a big bang singularity and ends in the de Sitter phase
$(\frac{\Gamma}{3},\frac{\Gamma^2}{3})$. }}
\end{figure}

\subsection{Bulk viscous cosmology:}

In  cosmology,  the  simplest  effective way to incorporate the bulk viscosity, is to use Eckart theory  \cite{Eckart}  (see  also  \cite{Zimdahl1, Zimdahl2}),  where  basically the  pressure $p$ is  replaced  by $p- 3 H \xi$, in which $\xi$ is the coefficient of bulk viscosity. 
Hence, the Friedmann equation and the Raychaudhuri's equations are modified as

\begin{eqnarray}
\rho &=& 3H^2,\label{bulk-friedmann1}\\
\dot{H}&=&- \frac{1}{2} (p+\rho)+ \frac{3}{2} H \xi.\label{bulk-friedmann2}
\end{eqnarray}
Now, using the barotropic EoS: $p= (\gamma- 1)\rho$, Raychaudhuri's equation can be written as

\begin{eqnarray}
\dot{H} &=& -\frac{3}{2} H^2 \gamma + \frac{3\xi}{2}H,
\end{eqnarray}
which coincides with (\ref{ray}) if one takes $\xi=\frac{\Gamma}{3}$. Therefore, our previous analysis holds in the case of a bulk viscous cosmology. In particular, the solution that
starts at $(0,0)$ and ends at the critical point $(\frac{\xi}{\gamma}, \frac{3\xi^2}{\gamma^2})$ is unphysical as we have already explained, and  contrary to the statement
of \cite{Chakraborty14}, it can not describe  a
scenario of  emergent universe. In connection with that, we note that the evolution of the universe driven by some bulk viscous pressure leads to some interesting consequences in presence of curvature, or anisotropy or an another fluid in the Friedmann equation \cite{JDB1, JDB2, JDB3}.


\section{Thermodynamic arguments}
\label{sec:thermo}

In this section, we shall extract the thermodynamical information of the present model. In principle, the macroscopic systems tend toward a thermodynamical equilibrium, which forms the basis of the second law of thermodynamics where the entropy, $S$ of an isolated system never decreases, that means, $\dot{S}\geq  0$,
and should be concave ($\ddot{S}<0$) in the last stage of approaching thermodynamic equilibrium \cite{MP13, pavon1} (note that the derivative could be with respect  any relevant variable, but here to simplify, we have chosen  the cosmic time).
In the FLRW universe, one may formulate this as follows: The entropy of the apparent horizon plus the matter or any fields enclosed by the horizon should satisfy $\dot{S}_h+ \dot{S}_{\gamma} \geq  0$, where $S_h$ stands for the entropy of the apparent horizon and $S_{\gamma}$ for the matter fields. On the other hand, $\ddot{S}_h+ \ddot{S}_{\gamma}  < 0$   at very late time, and positive at early times (see the discussion performed in Section II of \cite{MP13}). The entropy of the apparent horizon is defined as $S_h = k_{B} \mathcal{A}/4\, l_{pl}^2$, where $k_{B}$ is the Boltzmann's constant, $\mathcal{A}= 4 \pi r_h^2$, is the horizon area in which $r_h= \frac{1}{H}$ being the Hubble radius \cite{Bak}, and $l_{pl}=\sqrt{\frac{1}{8\pi}}$ is the Planck's length in the units used in the present work.




Now, taking the differentiation of $S_h$  with respect to the cosmic time, one gets
	\begin{eqnarray}\label{haro1new}
	\dot{S}_h & =-\frac{2\pi k_B}{l_{pl}^2}\frac{\dot{H}}{H^3}=\frac{24\pi^2 k_B\gamma}{H}\left(1-\frac{\Gamma}{3H}\right),
	\end{eqnarray}
	which shows that $\dot{S}_h>0$ for $H>\frac{\Gamma}{3}$. In the same way one obtains
	\begin{eqnarray}
	\label{haro2new}
	\ddot{S}_h & =-{24\pi^2 k_B\gamma} \frac{\dot{H}}{H^2}\left(1-\frac{2\Gamma}{3H}\right),
	\end{eqnarray}
	Now, since $\dot{H}<0$ for  $H>\frac{\Gamma}{3}$, hence ${S}_h$ is convex for $H>\frac{2\Gamma}{3}$ and concave for
	$\frac{2\Gamma}{3}>H>\frac{\Gamma}{3}$.

Now, considering the fluid, we recall the Gibb's equation $T dS_{\gamma}= d (\rho\, V) + p dV$, where $V= 4\pi r_h^3/ 3$, and using the cosmic time a simple calculation leads to





\begin{align}\label{haro3new}
	T\dot{S}_{\gamma} &=6\gamma\pi\left(1-\frac{\Gamma}{3H}\right)(3\gamma-2).
	\end{align}
Then, since standard matter satisfies $1\leq \gamma\leq 2$, so one may conclude as in case of $S_h$ that,
	$\dot{S}_{\gamma}>0$ is always true for $H>\frac{\Gamma}{3}$.
	To calculate the second derivative,
	we will use the equation \cite{Zimdahl2, LB14}
	\begin{eqnarray}
	\frac{\dot{T}}{T}=	(\Gamma-3H)\frac{\partial p}{\partial \rho},
	\end{eqnarray}
	that from the Equation of State $p=(\gamma-1)\rho$ and  using
	equation (\ref{ray}) could be written as follows
	\begin{eqnarray}
	\frac{\dot{T}}{T}=\frac{2(\gamma-1)}{\gamma}\frac{\dot{H}}{H}\Longrightarrow
	T=T_0\left( \frac{H}{H_0} \right)^{2\left(\frac{\gamma-1}{\gamma}\right)}	.
	\end{eqnarray}
	
	Then, a simple calculation shows that the second derivative of $S_{\gamma}$ is of the order
	\begin{align}\label{haro4new}
	\ddot{S}_{\gamma} & \sim -\frac{\dot{H}}{H^{\frac{3\gamma-2}{\gamma}}} \left(1-\frac{3\gamma-2}{6(\gamma-1)}\frac{\Gamma}{H}\right),
	\end{align}
	meaning that ${S}_{\gamma}$ is convex for $H>\frac{3\gamma-2}{6(\gamma-1)}\Gamma$, and concave for
	$\frac{3\gamma-2}{6(\gamma-1)}\Gamma>H>\frac{\Gamma}{3}$.

From that results one can conclude that
	$$\dot{S}_h+\dot{S}_{\gamma}\geq 0,$$
	from the big bang singularity to the de Sitter regime  given by the fixed point $H=\frac{\Gamma}{3}$. Moreover, since for $1<\gamma\leq 2$ one has $\frac{3\gamma-2}{6(\gamma-1)}\geq \frac{2}{3}$, then
	$({S}_h+{S}_{\gamma})$ is convex for $H>\frac{3\gamma-2}{6(\gamma-1)}\Gamma$ (if it was concave, the Universe could have reached the thermodynamical equilibrium before entering
in the stable de Sitter regime),  and concave for $\frac{2\Gamma}{3}>H>\frac{\Gamma}{3}$,  that is, the Universe eventually goes to the thermodynamical equilibrium stage characterized by a stable, and thus, never-ending, de Sitter regime with $H_{\infty}=\frac{\Gamma}{3}$.  One may note that for $\gamma = 1$, $T= T_0 =$ constant, thus it is readily seen that $\ddot{S}_{\gamma} < 0$ since $\dot{H}< 0$, that means for $H> \Gamma/3$.

Finally,
 when quantum corrections to Bekenstein-Hawking entropy law are encountered, the
entropy of black hole horizons is generalized into \cite{MP13}

\begin{align}\label{spnew1}
S_h & = k_B \left[ \frac{\mathcal{A}}{4 l_{pl}^2} - \frac{1}{2} \ln \left(\frac{\mathcal{A}}{l_{pl}^2} \right)\right],
\end{align}
in addition to that we have some higher order corrections given in \cite{new1, new2}. However, assuming this definition applies to the cosmic apparent horizon \cite{MP13}, one may find the modifications due to the correction term.
It is easy to find that
\begin{eqnarray}
\dot{S}_h=-\frac{3\gamma k_B}{2l_{pl}^2}\left(H-\frac{\Gamma}{3}\right)\left( -\frac{2\pi}{H^2}+l_{pl}^2 \right),
\end{eqnarray}
which is positive for $\frac{1}{H}>\frac{l_{pl}}{\sqrt{2\pi}}$, that is, it is an increasing function for all the values of the Hubble radius greater than the Planck length, i.e., when the classical picture of our Universe is allowed, one has  $\dot{S}_h>0$. In the same way one has
\begin{eqnarray}
\ddot{S}_h=-\frac{3\gamma \dot{H}}{2l_{pl}^2}\left[l_{pl}^2+\frac{2\pi}{H^2}\left(1-\frac{2\Gamma}{3H}\right)    \right].
\end{eqnarray}

Since $\dot{H}$ is negative, one can see that for large values of $H$ the function $S_h$ is convex, and for values near the fixed point $\frac{\Gamma}{3}$, it is concave, because one has
\begin{eqnarray}
\ddot{S}_h\left(\frac{\Gamma}{3}\right)=-\frac{3\gamma \dot{H}}{2l_{pl}^2}\left[l_{pl}^2-\frac{18\pi}{\Gamma^2}    \right],
\end{eqnarray}
which is negative due to the fact that $\Gamma$ is of the same order than the current value of the Hubble parameter.



\section{Extension to Loop Quantum Cosmology}
\label{sec:lqg}

For open systems with particle creation governed by the equation $\dot{N}= \Gamma N$, the Friedmann, 
conservation and the Raychauduri equations in Loop Quantum Cosmology (LQC) respectively take the forms  (see \cite{LQC} for a review)

\begin{eqnarray}
H^2&=& \frac{\rho}{3} \left(1-\frac{\rho}{\rho_c}\right),\label{lqg-friedmann1}\\
\dot{\rho}&=& - 3 H \left(1-\frac{\Gamma}{3H}\right) (p+ \rho),\label{lqg-conservation}\\
\dot{H}&=&-\frac{1}{2}\left(1-\frac{\Gamma}{3H}\right) (p+ \rho)\left(1-\frac{2\rho}{\rho_c}\right)\label{lqg-raychauduri}
\end{eqnarray}
where $\rho_c$ is the ``critical density'' (the energy density at which the universe starts bouncing). Note that the holonomy corrected Friedmann equation depicts an ellipse in the phase space $(H,\rho)$, and when $\Gamma$ is constant, the dynamical system becomes singular at the bouncing point $(0,\rho_c)$ because
the Raychauduri equation diverges at $(0,\rho_c)$, then to have a non-singular bounce, which is one of the advantages of LQC, one has to assume that $\Gamma$ changes  at different scales vanishing at the bouncing point. Since near the bounce, $H$ decays as $\sqrt{1-\frac{\rho}{\rho_c}}$, in order to have a non-singular bounce, it seems natural to choice $\Gamma$ as follows

\begin{eqnarray}\label{rate-varying}
\Gamma=\Gamma_0\left(1-\frac{\rho}{\rho_c}\right)^{\alpha+\frac{1}{2}}\quad \mbox{with} \quad \alpha\geq 0,
\end{eqnarray}
and then, in that case,  assuming that $\Gamma^2_0\ll \rho_c$ (recall that, $\Gamma_0$ is of the order of the current value of the Hubble parameter and $\rho_c\cong 0.4 \rho_{pl}$ \cite{Meissner} being $\rho_{pl}$ the Planck's energy density),
for a fluid with EoS $p=(\gamma-1)\rho$,  with $\gamma>0$,
the dynamics that goes from $(0,~0)$ to $\left(\frac{\Gamma_0}{3},\frac{\rho_c}{2}\left(1-\sqrt{1-\frac{4\Gamma^2_0}{3\rho_c}}\right)\right)
\cong\left(\frac{\Gamma_0}{3},\frac{\Gamma^2_0}{3}\right) $,
depicts, in a clockwise direction, a bouncing universe that starts in the contracting phase, bounces
at $(0,~\rho_c)$, and ends in the de Sitter phase $\left(\frac{\Gamma_0}{3},\frac{\rho_c}{2}\left(1-\sqrt{1-\frac{4\Gamma^2_0}{3\rho_c}}\right)\right)$ (see figure 2).
Also, note that, the dynamics from $(0,~0)$ to $\left(\frac{\Gamma_0}{3},\frac{\rho_c}{2}\left(1-\sqrt{1-\frac{4\Gamma^2_0}{3\rho_c}}\right)\right)$ in the
anticlockwise direction is un-physical as discussed in the previous section.}

On the other hand, using (\ref{EoS}),
in LQC the effective EoS parameter is given by

\begin{eqnarray}
w_{eff}= -1+ \gamma \left(1-\frac{\Gamma_0}{{3H}}\left(1-\frac{\rho}{\rho_c}\right)^{\alpha+\frac{1}{2}}\right),
\end{eqnarray}
which shows that initially $w_{eff}=\infty$,  and when  $\rho\gg \Gamma_0^2$,
one has $w_{eff}\cong -1+\gamma$ (which includes the bounce), and finally, at late times ($\rho\sim \Gamma_0^2$), $w_{eff}\gtrsim -1$, that depicts the current accelerated expansion of the universe.

\begin{figure}[h]
\begin{center}
\includegraphics[scale=0.40]{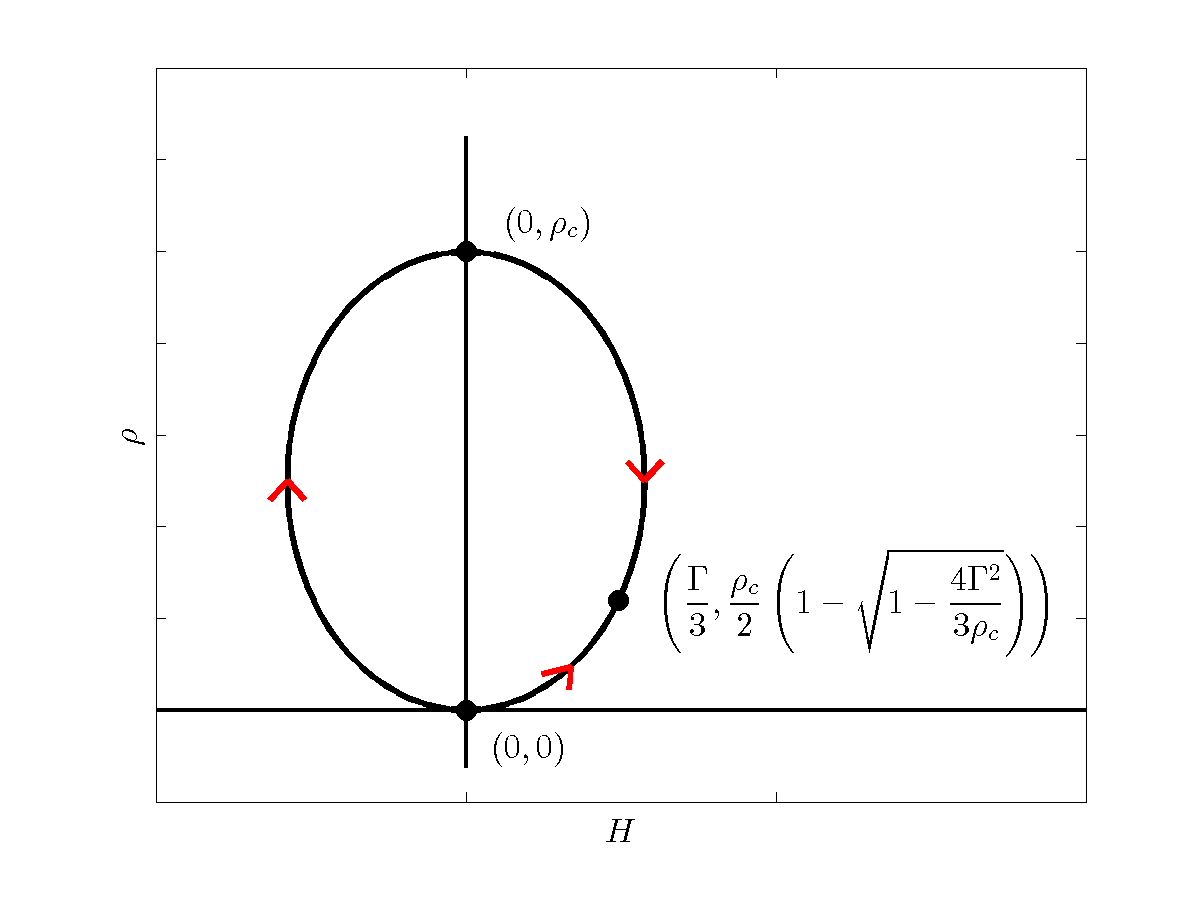}
\end{center}

\caption{{\protect\small The dynamics depicts a bouncing universe starting at $(0,0)$ and moving in the contracting phase towards $(0, \rho_c)$, where it bounces and enters into the final de Sitter regime.
}}
\end{figure}

Moreover,
if one chooses $\gamma=1-\epsilon$ with $0<\epsilon\ll 1$, i.e., one considers a nearly pressureless fluid, at early times one obtains {{}
\begin{eqnarray}
w_{eff}=-\epsilon-(1-\epsilon)\frac{\Gamma_0}{\sqrt{3\rho}}\cong -\epsilon-\frac{\Gamma_0}{\sqrt{3\rho}}
\cong -\epsilon-\frac{\Gamma_0}{{3H}}.
\end{eqnarray}}

Then, for a single scalar field, which mimics the fluid with EoS $p=-\epsilon \rho$, and thus,  drives the background of the Matter Bounce Scenario in LQC
{{} \cite{ewing}}. It has been recently showed
that the spectral index of {{} scalar} cosmological perturbations is given by {{} \cite{wilson1,eho}}
\begin{eqnarray}
 n_s-1=12w_{eff}=-12\epsilon-\frac{4\Gamma_0}{H},
\end{eqnarray}
which means that for modes that leave in the contracting phase, the Hubble radius satisfies $\frac{\Gamma_0}{\epsilon}\ll|H|\ll \sqrt{\rho_c}$ and for $\rho\ll \rho_c $ (at
this stage holonomy corrections could
be disregarded) one has
\begin{eqnarray}
 n_s-1=12w_{eff}\cong -12\epsilon,
\end{eqnarray}
which fits well with recent observational data $n_s-1=-0.0397\pm 0.0073$ \cite{Ade},  if one chooses $\epsilon\cong 0.0033$.
Now, once one has obtained the spectral index one can calculate its running, which is given by
\begin{eqnarray}
 \alpha_s\equiv \frac{\dot{n}_s}{\frac{d(\ln |aH|)}{dt}}=12\dot{w}_{eff}\left(\frac{H}{H^2+\dot{H}}\right).
\end{eqnarray}

During the matter domination \textbf{ $\frac{\Gamma_0}{\epsilon}\ll|H|\ll \sqrt{\rho_c}$,} one has $H^2+\dot{H}\cong -\frac{H^2}{2}$ and \textbf{$\dot{w}_{eff}\cong -\frac{1}{2}\Gamma_0$,} then one gets {{}
\begin{eqnarray}
 \alpha_s\cong - \frac{12 \Gamma_0}{H}.
\end{eqnarray}}

This running is negative, and
it belongs to the marginalized 95\% Confidence Level
(recent  Planck's data   states that $\alpha_s= -0.0134\pm 0.0090$ \cite{Ade})
for modes that leave in the contracting phase, when the Hubble radius belongs to the interval $10^3\,\Gamma_0\lesssim|H|\ll \sqrt{\rho_c}$.

\section{Summary and Discussion}
\label{discu}

Cosmological models powered by the gravitationally induced `adiabatic' particle creation have intensively been investigated in the FLRW universe as a possible source for late cosmic acceleration
\cite{SSL09, LJO10, LB10, LBC12, JOBL11, LGPB14, CPS14, NP15, FPP14, NP2016, RCN2016}. In connection with different particle creation rates, the possibility of constant particle creation rate has also been investigated in \cite{Prigogine98, Chakraborty14} where both the works claimed that the big bang singularity disappears, but in general which is not true. Thus, in order to clarify this issue, in the present work, we have considered a cosmological model in the flat FLRW universe driven by some constant particle creation rate, $\Gamma$. We found that the present model can be analytically solved which predicts that, at  very
early time, there was a big bang singularity in contrary to the results in \cite{Prigogine98, Chakraborty14}, and additionally, at late-time, our universe asymptotically approaches to the  de Sitter phase. We then consider the dynamical system consisting
of the Friedmann equation and the Raychaudhuri equation, which predicts that the system has only two critical points $(0,~0)$, $\left(\frac{\Gamma}{3},~\frac{\Gamma^2}{3}\right)$. The dynamics of
the universe along the parabolic path described by the Friedmann equation (\ref{friedmann1}) has been studied, which shows that the movement of the universe from $(0,~0)$ to
$\left(\frac{\Gamma}{3},~\frac{\Gamma^2}{3}\right)$ in the clockwise direction (see figure 1) is governed by the phantom fluid which totally unphysical due to absence of the radiation
and the matter dominated eras as predicted by the standard cosmology in
agreement with the observations. On the other hand, we found that, the movement of the universe towards the point  $\left(\frac{\Gamma}{3},~\frac{\Gamma^2}{3}\right)$
(see figure 1) starts from a big bang singularity and ends in the de Sitter phase asymptotically. Furthermore, we perform the same dynamical analysis for the universe
if it is dominated by some bulk viscous pressure. This analysis contradicts the existence of an emergent universe scenario driven by some constant bulk viscous pressure as discussed in \cite{Chakraborty14}. 
Moreover, we have shown that the present cosmological model driven by such constant creation rate approaches toward a thermodynamic equilibrium state in the late de Sitter phase. Finally, for the first time, we extend this adiabatic creation mechanism in the Loop Quantum Cosmology which is a promising candidate to understand the early physics of our universe since as in inflation it also provides a nearly flat power spectrum for cosmological perturbations. We found that in order to have a nonsingular bounce the creation rate should not be constant, rather it must depend on the energy scale (see equation (\ref{rate-varying})). The dynamics of the universe has been graphically shown in figure 2 where the dynamics from  $(0,~0)$ to $\left(\frac{\Gamma_0}{3},\frac{\rho_c}{2}\left(1-\sqrt{1-\frac{4\Gamma^2_0}{3\rho_c}}\right)\right)$ (see section \ref{sec:lqg}) in an anticlockwise direction is unphysical since it is governed by the phantom fluid and there are no such radiation and matter dominated eras. On the other hand, the dynamics from $(0,~0)$ to $\left(\frac{\Gamma_0}{3},\frac{\rho_c}{2}\left(1-\sqrt{1-\frac{4\Gamma^2_0}{3\rho_c}}\right)\right)$  in the clockwise direction represents a bouncing universe that starts in the contracting phase, bounces at $(0, \rho_c)$ - the critical point staying between the initial and the final points, and finally ends in the de Sitter regime executing the current accelerating universe. So, in the LQC frame the scale dependent matter creation rate could replace the big bang singularity by the big bounce one. Moreover, one obtains the inflationary parameters such as the spectral index, its running, which are in well agreement with the latest observational data \cite{Ade}.


\section{Acknowledgments}  We would like to thank Professor Jaume Amor\'os for his help in setting the figures.
This investigation of J. Haro has been supported
in part
by MINECO (Spain), project MTM2014-52402-C3-1-P. S. P acknowledges Science and Engineering Research Board (SERB), Govt. of India, for awarding National Post-Doctoral Fellowship (File No: PDF/2015/000640). We also  thank Rafael C Nunes for fruitful discussions. Finally, the authors are very grateful to the anonymous reviewers for their constructive suggestions which results in an improved version of the manuscript both in quality as well as in presentation.


\end{document}